\documentclass{llncs}

\usepackage{times}  
\usepackage{helvet}  
\usepackage{courier}  
\usepackage{url}  
\usepackage{graphicx}  

\usepackage{balance}       
\usepackage{graphics}      
\usepackage[T1]{fontenc}   
\usepackage{txfonts}
\usepackage{mathptmx}
\usepackage{color}
\usepackage{booktabs}
\usepackage{textcomp}
\usepackage{subfigure} 

\usepackage{microtype}        

\usepackage{todonotes}

\usepackage{xcolor}
\usepackage{xspace}

\newcommand\qt{\texttt{CityPulse}\xspace}

\begin{document}

\title{City of the People, for the People: Sensing Urban Dynamics via Social Media Interactions}
\subtitle{(Extended version of SocInfo'18 paper)}

\author{Sofiane Abbar \inst{1} \and Tahar Zanouda \inst{1} \and Noora Al-Emadi \inst{1} \and Rachida Zegour \inst{2}}

\authorrunning{S. Abbar et al.} 
\tocauthor{Sofiane Abbar, Tahar Zanouda, Noora Al-Emadi, Rachida Zegour}

\institute{Qatar Computing Research Institute, HBKU, Doha, Qatar\\
\and
Abderrahmane Mira - B\'eja\"ia University, Algeria\\
\email{\{sabbar, tzanouda, nalemadi\}@qf.org.qa, zegourr@gmail.com}
}

\maketitle

\begin{abstract}
Understanding the spatio-temporal dynamics of cities is in the heart of many applications including urban planning, zoning, and real-estate construction.
So far, much of our understanding about urban dynamics came from traditional surveys conducted by persons or by leveraging 
mobile data in the form of Call Detailed Records.
However, the high financial and human cost associated with these methods make the data availability very limited. 
In this paper, we investigate the use of large scale and publicly available user contributed content, in the form of social media posts to understand the urban dynamics of cities. We build activity time series for different cities, and different neighborhoods within the same city to identify the different dynamic patterns taking place. Next, we conduct a cluster analysis on the time series to understand the spatial distribution of patterns in the city. Our instantiation for the two cities of London and Doha shows the effectiveness of our method.  
\end{abstract}



\section{Introduction}
\label{sec:intro}

After the revolutionary growth of the modern industry in the late 18th century, cities have become a niche for new opportunities that attracted huge numbers of local and international migrants. In 2015, we were 53.89\% to live in urban areas and this number is projected to reach 70\% by the end of 2050~\cite{unhabitat}. This fast urbanization has exposed different services and infrastructures in cities to an increasing stress, leading to a plethora of urban challenges related to human mobility (e.g., traffic congestion), public health (e.g, air pollution), neighborhood deprivation (e.g., lack of parks), and economical difficulties (e.g., unemployment). The situation is such that the United Nation has declared managing urban areas as one of the most important development challenges of the 21st century. Thus, understanding different patterns of urban dynamics taking place in our cities is of a paramount importance for many applications.

Until now, much of our understanding about urban dynamics came from traditional surveys conducted by human agents either physically or by phones \cite{morenoff2001neighborhood,theobald2001land}. While this way of collecting data provides detailed information about urban behaviors,  it remains hard to update and present many weaknesses regarding generalization and scalability.
More recently, researchers have looked into the use of mobile data such as Call Detailed Records (CDRs) to model urban dynamics \cite{Reades:2007,gonzalez2008understanding,Grauwin2015}. These attempts have been quite successful in building accurate models for different dynamics such as mobility, traffic congestion, and land use at city-wide scales. However, these two methods suffer major drawbacks related to the high financial (CDRs) and labor (human operated surveys) costs required to acquire the data needed, which makes its availability very limited. For instance, CDRs are not only expensive, but are also highly proprietary (telecommunication companies) posing serious data sharing challenges. 

Alternatively, social media is notorious for providing its users with a means of documenting the
minutiae of their daily lives, including places they go and activities they engage with \cite{cranshaw2012livehoods,steiger2015twitter}. 
In this paper, we investigate the use of large scale and widely available public user generated data, in the form of social media posts, to understand urban dynamics taking place in cities and neighborhoods. We are particularly interested in data that is geo-located, i.e. associated with an accurate pair of latitude and longitude coordinates, that is generated by users on different social media platforms such as Twitter, Foursquare, and Instagram. 
We argue that analyzing geo-located posts can provide an interesting angle to look at the city as a holistic dynamic system. Our analysis is driven by the following two research questions.

\textbf{RQ 1. Can we use social media -- and Twitter in particular -- to get insights into urban dynamics in different cities?}
Inspired by th work of Reades et al. \cite{Reades:2007} on exploring the use of mobile data, we hypothesize that spaces can be characterized by the volume of social media use over time. By analyzing the use of social signatures, we can show how social media interactions are related to urban activities. 
After all, posting a check-in in a restaurant, at a stadium, or from the college amphitheater to inform friends and the world about our real-time location and activity has become so common in the era of social media . 
This unprecedented willingness to share with others provides a novel view about urban spaces as seen through social media activity. It also gives good insights about the spatio-temporal dynamics of urban life, leading to a better understanding of the city's complex hourly, daily, and weekly rhythms.

In this study, we collected geo-located tweets over a period of two months from two different cities: London (UK) as a mature and developed city, and Doha (Qatar) as a fast-growing multi-cultural city. The data is transformed into time-series that capture the volume of social media activity over time, which are then normalized to form typical weekly signatures. These signatures are analyzed to discover common weekly patterns in which more activity is observed during working days compared to weekends, and daily cycles that reveal activities such as waking-up in the morning, going to work, then heading back home. A decomposition of the typical signature time-series allows to disentangle general weekly trends from residuals that capture special events such as holidays.

\textbf{RQ 2. Is social media data sensitive enough to pick out more fine grained variations at the level of city neighborhoods?}
To answer this question, we perform a cluster analysis on typical weekly signatures in the format of time-series created for different neighborhoods. To this end, a city is partitioned into different zones corresponding to a particular administrative division such counties (in the US), boroughs (in UK), and districts (in Doha). The objective is to discover the different patterns and urban rhythms that characterize different areas of the city presenting different cultural, social and economic properties. The cluster analysis allows also to spot neighborhoods that have similar patterns. As a clustering algorithm, we use $k-means$ method that aims at maximizing the similarity among items belonging to the same cluster, and maximizing the dissimilarity among items belonging to different clusters. As the clustering is operated upon time-series, we use the Dynamic Time Warping (DTW) distance~\cite{dtw} that has been shown to outperform all other distances when comparing time-series~\cite{wang2013experimental}. Interestingly enough, when the cluster analysis is performed on the zones of both cities together, it identified neighborhoods that share similar spatio-temporal patterns across the two cities.

\textbf{Positioning.} Using social media data to characterize urban dynamics in neighborhoods is not new \cite{noulas2011exploiting,cranshaw2012livehoods,steiger2015twitter,ArribasBel2015,krumm2017tweetcount}. Therefore, we analyze in the following major works done in this area and try to contrast them to our work.

In a study by \cite{noulas2011exploiting}, authors looked at modeling human activity and geographical areas via spectral clustering of Foursquare data. The main idea is to split a region into equally sized rectangles, and then characterize each rectangle with a vector containing the total number Foursquare places belonging to different categories. While this approach is interesting, it completely dismisses the temporal aspect of human activity and focuses solely on the geographical distribution and popularity of places (e.g. shops, restaurants, schools, etc.) Our framework's main contribution is to allow a spatio-temporal analysis of human dynamics in cities. 
The {\it Livehoods Project} is another influential work in this area \cite{cranshaw2012livehoods}. It aims at using Foursquare data for explaining dynamics in cities, and use for that spectral clustering on types of Foursquare venues present in different areas of a city. However, similar the work by \cite{noulas2011exploiting}, this project ignores the temporal aspect in the data, which we believe is very important in modeling human dynamics. Indeed, two areas with similar facilities may show different temporal behaviors (e.g. people may stay late at night in one, but not in another.)
More recently, \cite{krumm2017tweetcount} demonstrated that using tweet counts could help identifying the land use profile of a neighborhood in a city. Authors used temporal features such as average number of tweets observed within different hours of the day to train a classifiers to label neighborhoods. The main weakness of this approach resides in the many heuristics introduced in defining land use profiles and selecting the ones to test against. Indeed, while land use has already a well-known classification into commercial, business, residential, industrial, etc. Authors provide their own definition of land use profiles on the only basis of resident and business count. A typical profile in their case would be a one that represents a neighborhood (a square cell) with low number of residents and high number of businesses. Our work is different from this one in two aspects. First we introduce a clustering approach using dynamic time warping on weekly time-series. Second, we study well-defined neighborhoods (represented as polygons) that correspond to actual administrative zones.    

In order to demonstrate the effectiveness of using social media based time-series to capture the different urban rhythms within cities, we built and released a tool that we call \qt\footnote{http://citypulse.qcri.org}. The main objective of \qt is to allow a visual inspection of the different spatio-temporal patterns captured in different cities. 

Subsequently, we first review related work that use rich datasets to characterize urban dynamics.
Next, we introduce the data gathering and pre-processing steps followed by discussions of some preliminary insights from the data. 
Then, we present the clustering algorithm used to cluster time series along side with the distance used. 
Next, we comment on the clustering outcomes for both inter-city and intra-city clusterings. 
Finally, Section \ref{sec:conclusion} concludes the paper with final remarks and future directions.

\section{Related Work}
\label{sec:relatedwork}

We review in this section related literature concerned with exploring big urban data for modeling and understanding urban dynamics.

\subsection{Using Mobile Data for Urban Analytics}
The wide-spread and adoption of mobile phone technologies have allowed telecommunication companies to gather massive data sets about the spatio-temporal daily activities of people that yield better understanding of how our cities function \cite{batty2007cities}. These rich mobile phone datasets have unlocked the potential for several urban related applications such as traffic congestion \cite{ccolak2016understanding}, human mobility patterns~\cite{gonzalez2008understanding}, exposure to air pollution \cite{nyhan2016exposure}, and sensing urban dynamics~\cite{ratti2006mobile} to name but a few.  
One of the closest work to ours is the one done by MIT's Senseable City Lab\footnote{http://senseable.mit.edu/} in which they partnered with different telecommunication companies based in Europe and used the call data records (CDRs) to profile cities and neighborhoods using typical weekly signatures (TWS) in the format of time-series. Reades et al.~\cite{Reades:2007} present a nice overview of their related urban computing projects in. Following the same line of research, Grauwin et al.~\cite{Grauwin2015} proposed a framework for comparative science of cities by comparing the spatio-temporal dynamics of neighborhoods, represented as time-series featuring different mobile phone activities such as: calls, messages, and data usage. Toole et al.~\cite{toole2012inferring} propose also to characterize neighborhoods using mobile data based time-series that are used to classify land use of different neighborhoods in Boston. The main finding is that time-series of mobile activity can be used to figure out what type of urban activities are taking place in different areas by inferring the land use of those areas. Examples of land use classes considered in the study are: residential, commercial and industrial. 

\subsection{Social Media, and Urban Studies}
While the aforementioned studies have enabled an unprecedented city-wide understanding of different urban phenomenons, they suffer a serious drawback in that the data they rely on, i.e., mobile phone data, are highly proprietary, expensive, and have many other challenges that limit their availability. This is why we propose in this work to explore other sources of data, more affordable and widely available, such as social media posts, to evaluate their suitability for sensing urban dynamics.

In addition to the works mentioned in the introduction \cite{noulas2011exploiting,cranshaw2012livehoods,steiger2015twitter,ArribasBel2015,krumm2017tweetcount}, there are some on-going efforts within the HCI community that aim at understanding the role of data in city operations and urban phenomena in general. For instance, The work of Fabio et al.~\cite{miranda2016urban} aimed at designing a new  a visual exploration framework which allows urban planners
to explore the pulses within and across multiple cities under different conditions. Moreover, the tool can capture the spatio-temporal activity in a city across multiple temporal resolutions.  Aoki et al.~\cite{aoki2009vehicle} presented a qualitative analysis of the landscape of environmental action. Authors have built an environmental air quality sensing system and deployed it on street sweeping vehicles in San Francisco to serve as "research vehicle" that generated different types of data.
Blom et al.~\cite{blom2010fear} proposed a mobile communication system that helps alleviate fear and anxiety experienced in   the urban context, particularly by women, in three different cities. 
More recently, McMillan et al.~\cite{McMillan:2016} investigated the role and impact of data availability on urban initiatives conducted by different cities. 
Examples of using geo-located Twitter data include works from  Bassolas et al.~\cite{bassolas2016touristic} to measure touristic site attractiveness through the lenses of Twitter. Although, Abbar et al.~\cite{abbar2015you}  correlated mentions of food in Twitter with prevalences of diabetes and obesity in different urban areas in the US. Another interesting work is done by Liccardi et al.~\cite{liccardi2016know} in which they have collected and labeled geo-located tweets and then asked 45 participants to analyze how accurately they could infer the functional location of the original data owners. 
We capitalize on these previous studies and try to expand the use of geo-located social data in a more holistic approach to better capture urban dynamics. 

\section{Data Gathering and Processing}
\label{sec:data}

We discuss in this section the geographical context of the study as well as the data gathering processes we used to collect and clean geo-tagged data from Twitter.

\subsection{Geographical Context}
\begin{figure}[!h]
\centering
\subfigure[Doha population]{\includegraphics[width=0.26\columnwidth]{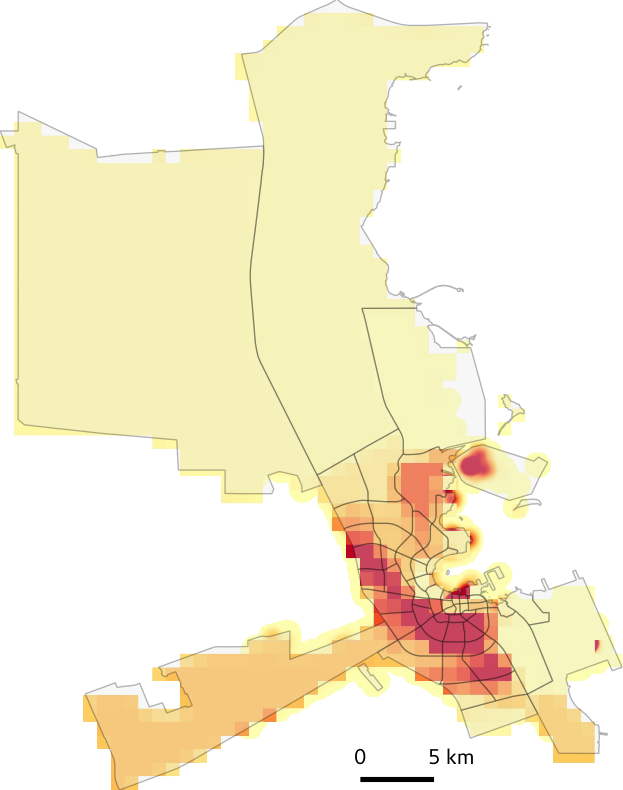}}
\subfigure[London population]{ \includegraphics[width=0.42\columnwidth]{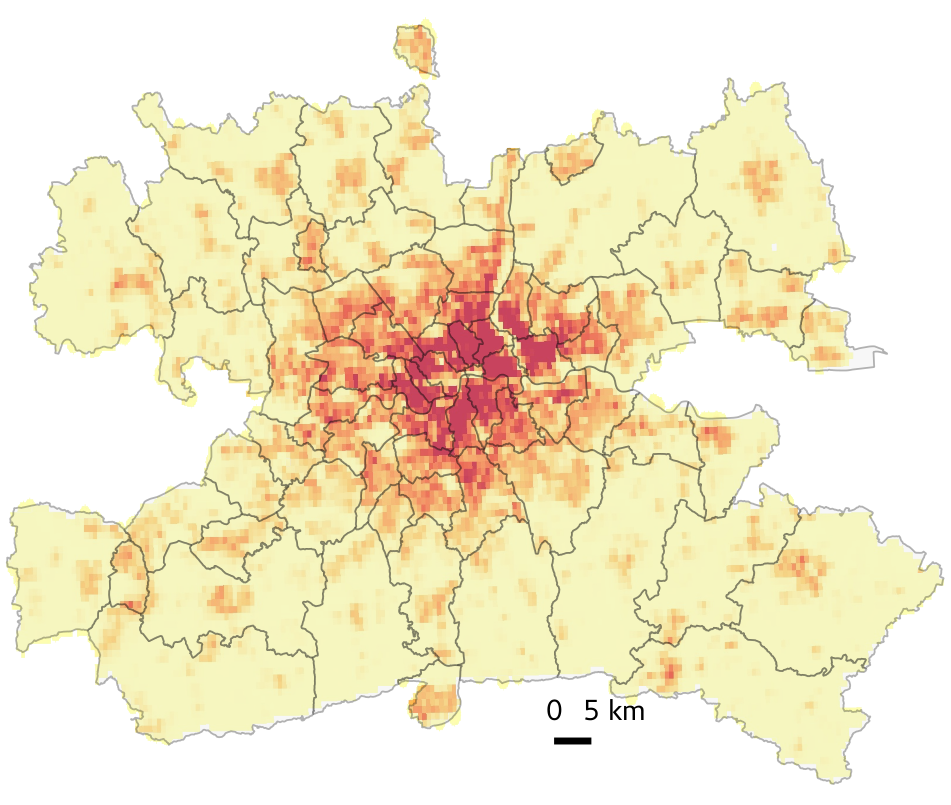}}\\
\subfigure[Doha activity]{\includegraphics[width=0.26\columnwidth]{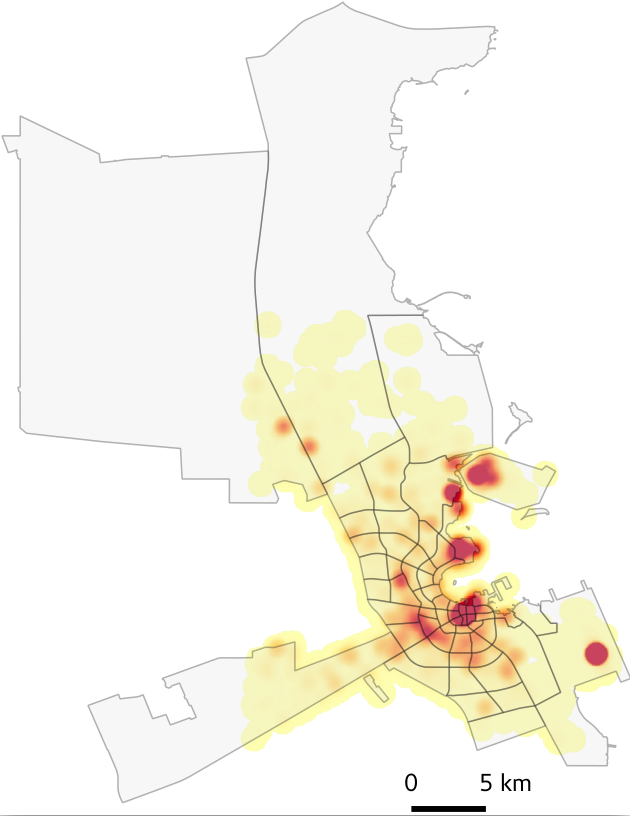}}
\subfigure[London activity]{ \includegraphics[width=0.42\columnwidth]{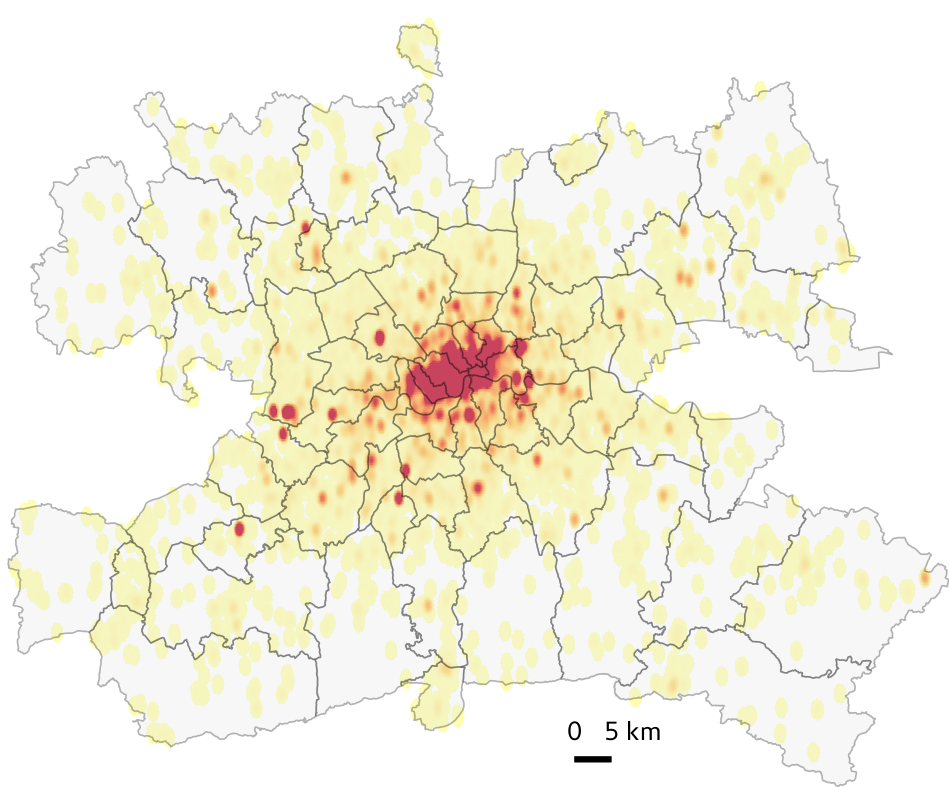}}

\caption{Administrative maps of Doha and London (or more accurately Extended Great London). The two panels (a) and (b) visualize the spatial distribution of population densities in the two cities of Doha and London respectively. Panels (c) and (d) showcase the spatial distribution of their recorded Twitter activity. Note the difference in scale for the maps of the the two cities.}
\label{fig:city_admin}
\end{figure}

Our study focuses on the two cities of Doha and London spotted in Figure \ref{fig:city_admin}. 
Doha is a fast growing city located in the middle east, with an unseen annual population growth of 10\%. Doha is also undertaking deep infrastructural changes related to the organization of FIFA 2022 Wold Cup. New community districts are being constructed in the periphery of the city and a whole new multi-modal transportation system composed of metro and rail lanes is being deployed. Doha is also home to a rich international mix of people. Indeed, locals only represent about 10\% of the total population estimated to 2.2M in the last 2015 census\footnote{https://data.worldbank.org/country/qatar}. The city of Doha is one of the seven municipalities in Qatar, and is divided into 61 different zones whose sizes increase as they move away from the central business district. 
London on the other hand is an example of a mature and developed city, home to more than 8.2M people. Extended Great London area encompasses 61 district community boroughs\footnote{https://data.worldbank.org/country/united-kingdom}. 

Population composition of the two cities and the different life styles observed in European cold London and middle-estern hot Doha are good motivations to study their different underlying urban dynamics.

\subsection{Twitter Data Collection}
For this study, we use the \textit{location} filter Twitter Streaming API\footnote{https://dev.twitter.com/streaming/overview} that allows to retrieve geo-located tweets matching a specified input region. The API takes as input a comma-separated list of longitude,latitude pairs describing a set of bounding boxes to filter Tweets. Only geo-located Tweets falling within the requested bounding boxes will be included. As per Twitter  documentation, each bounding box should be specified as a pair of longitude and latitude pairs, with the southwest corner of the bounding box coming first. We run the API for the two cities of London and Doha. For each city, we use Klokantech\footnote{http://boundingbox.klokantech.com/} service to request its rectangular bounding box. We then use administrative shapefile extracted from OpenStreetMap\footnote{http://openstreetmap.org} to assign geo-located tweets to the right administrative zone. Tweets that do not match any of the administrative zones are ignored. 
The data used in this study spans three months June--August 2017, and consists of 60,197 tweets posted from Doha and 152,007 tweets from London. Figure~\ref{fig:city_admin} gives a glimpse into the spatial distribution of geo-located data in the two cities. We clearly see that activity in London is concentrated around its central business district (CBD), whereas in Doha the activity is spread across the cost-line, with noticeable clusters at the airport (south east), the Souq area (center), and West-Bay (north east). 

For population data shown in Figures~\ref{fig:city_admin}(a) and (b), we use Gridded Population of the World (GPWv4) by NASA\cite{gpw}. The data consists of estimates of human population density based on counts consistent with national censuses and population registers, for the years 2000, 2005, 2010, 2015, and 2020. A proportional allocation gridding algorithm, utilizing approximately 12.5 million national and sub-national administrative units, is used to assign population values to approximately 1 km grid cells. While this data is widely recognized for its accuracy, especially for well-established developed cities such as London, the projections are often wrong for fast-growing cities such as Doha, which started a huge population growth in late 2011 just after they released the census report used in GPWv4\cite{gpw},  explaining the less obvious correlation between population and activity densities in this city.

\section{Preliminary Insights}
\label{sec:insights}

Now, we dig into data to retrieve some insights about the rhythms of the two cities as captured through the lenses of Twitter activity. In order to better capture the vibe of cities, we use the methodology presented in \cite{Reades:2007} to build their average or what can be coined as representative ``\textit{typical week signatures}'' (TWS).  
$TWS$ of a region is a 168 (24 hours $\times$ 7 days) long time-series reporting the hourly observed Twitter activity in the whole region. Each value in $TWS$ represents the average (typical) count (level) of activity observed in the region in similar hours of the same day on the week throughout the three month period of data. That is, the value in TWS for Tuesday 8am is the average of all observations that happened on all Tuesday at 8am in the collected data.   

\subsection{Insights at the Macro Level of Cities}
\begin{figure}[!h]
\centering
\includegraphics[width=0.79\columnwidth]{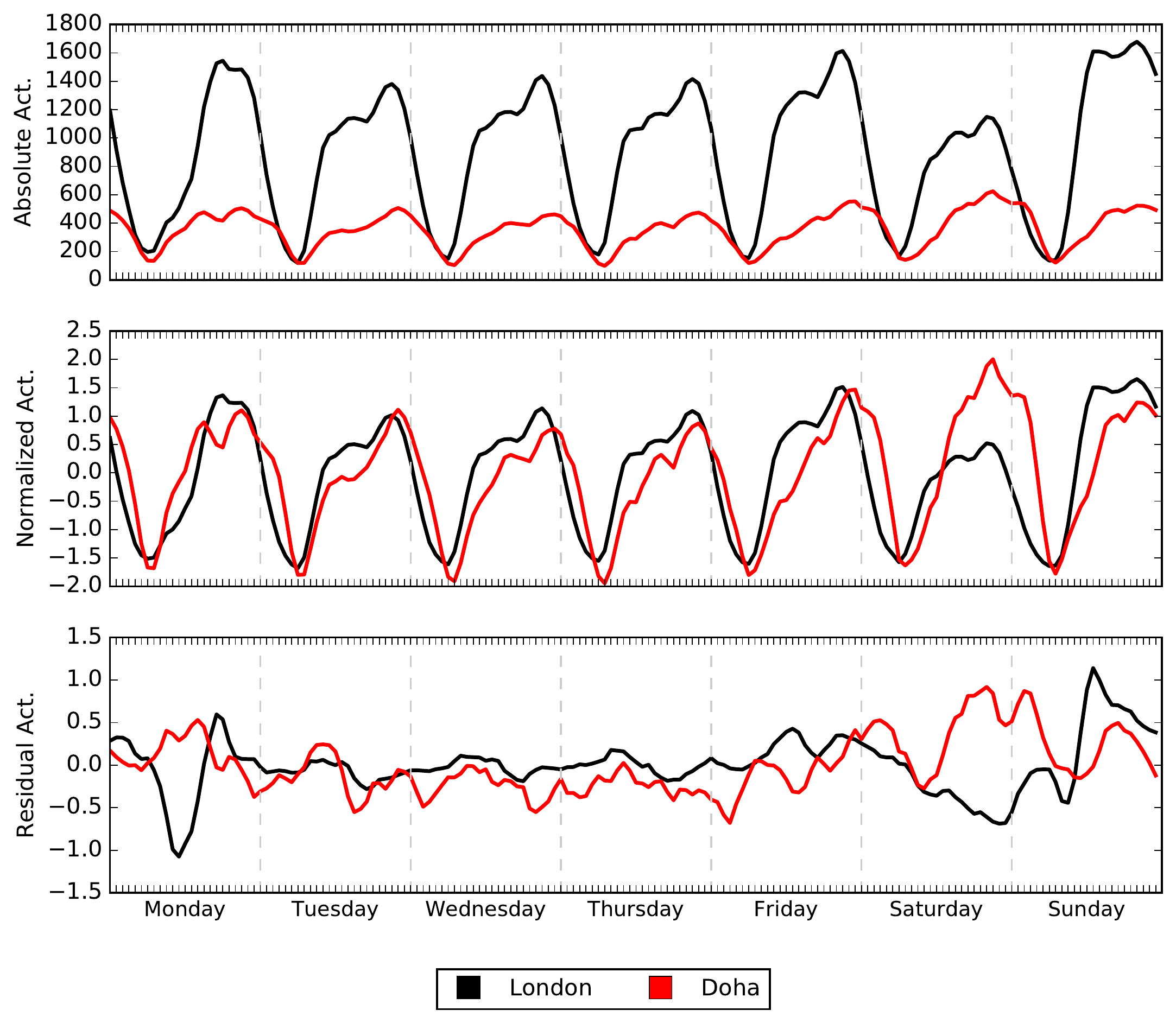}

\caption{Typical weekly signatures for Doha and London. The top panel plots the average absolute activity (number of tweets) typically observed at different hours of the week. The middle panel plots the $z-score$ normalized time-series of the two cities. The bottom panel shows residual activity capturing spacial events. The city of London shows a significant drop in activity on Saturdays whereas the activity in the city of Doha increases.}
\label{fig:comparative_signatures}
\end{figure}

We focus in this section on comparing the typical weekly signatures of the two cities of Doha and London.  
We plot in Figure \ref{fig:comparative_signatures} the two typical weekly signatures of Doha and London. As all timestamps in the data come in GMT timezone, Doha time series has been shifted by three hours to reflect its actual GMT+3 timezone. 
Examining absolute counts (top panel) for the two cities of Doha and London, reveals big discrepancies as the average activity level in the two cities differs significantly.
One could easily spot that London presents a surge in activity on Fridays night and a significant drop in activity on Saturdays (first day of the weekend.) Doha on the contrary, shows a relatively steady levels throughout the week, with a slight increasing tendency towards Saturdays (second day of the weekend in Doha.) In order to correctly compare for the population differences in the two cities, we normalize their TWS time-series using z-score. Normalized time-series have usually a zero mean and a unit standard deviation. Thus, the z-normalized time-series of a given time-series $TWS_{city}^{act}$ is calculated as follows: 
$TWS_{city}^{norm}(t) = \frac{TWS_{city}^{act}(t) - \mu_{TWS_{city}^{act}}}{\sigma_{TWS_{city}^{act}}} $

Normalized time-series of Doha and London are plotted in the second panel of Figure~\ref{fig:comparative_signatures}. These time-series show now somewhat similar patterns reflecting the typical circadian rhyme of a city in which people wake up in the morning, go to work, have dinner, and come back home. The max activity happens in both cities toward the end of the day. Interestingly enough, we see that activity in Doha is slightly shifted compared to that in London, in that the morning activity starts later and the last later at night. This is counter intuitive given for instance that official working hours for many administrations and schools in Doha are from 7am to 3pm whereas it is 8am-4pm in London. However, the city of Doha is notoriously known for its night activity, mainly due to the hot weather in the day. Indeed, it is common to see families arriving to public parks after 8pm and stay till midnight, even in weekdays.

The residual activity plots shown in the bottom panel of Figure~\ref{fig:comparative_signatures} reveal an interesting increasing tendency of activity in the city of Doha on Fridays and Saturdays which form the weekend in Qatar. On the contrary, the activity tends to decrease on the same days in the city of London, before it picks up again on Sunday. Residual activity can be interpreted as the deviation from expected of activity on a given day/hour. We compute residual activity by decomposing the $z$-normalized time-series into seasonal and residual. In short, residual time-series are obtained by subtracting the seasonal time-series from the initial normalized time-series.  

\begin{figure}[!h]
\centering
\includegraphics[width=0.78\columnwidth]{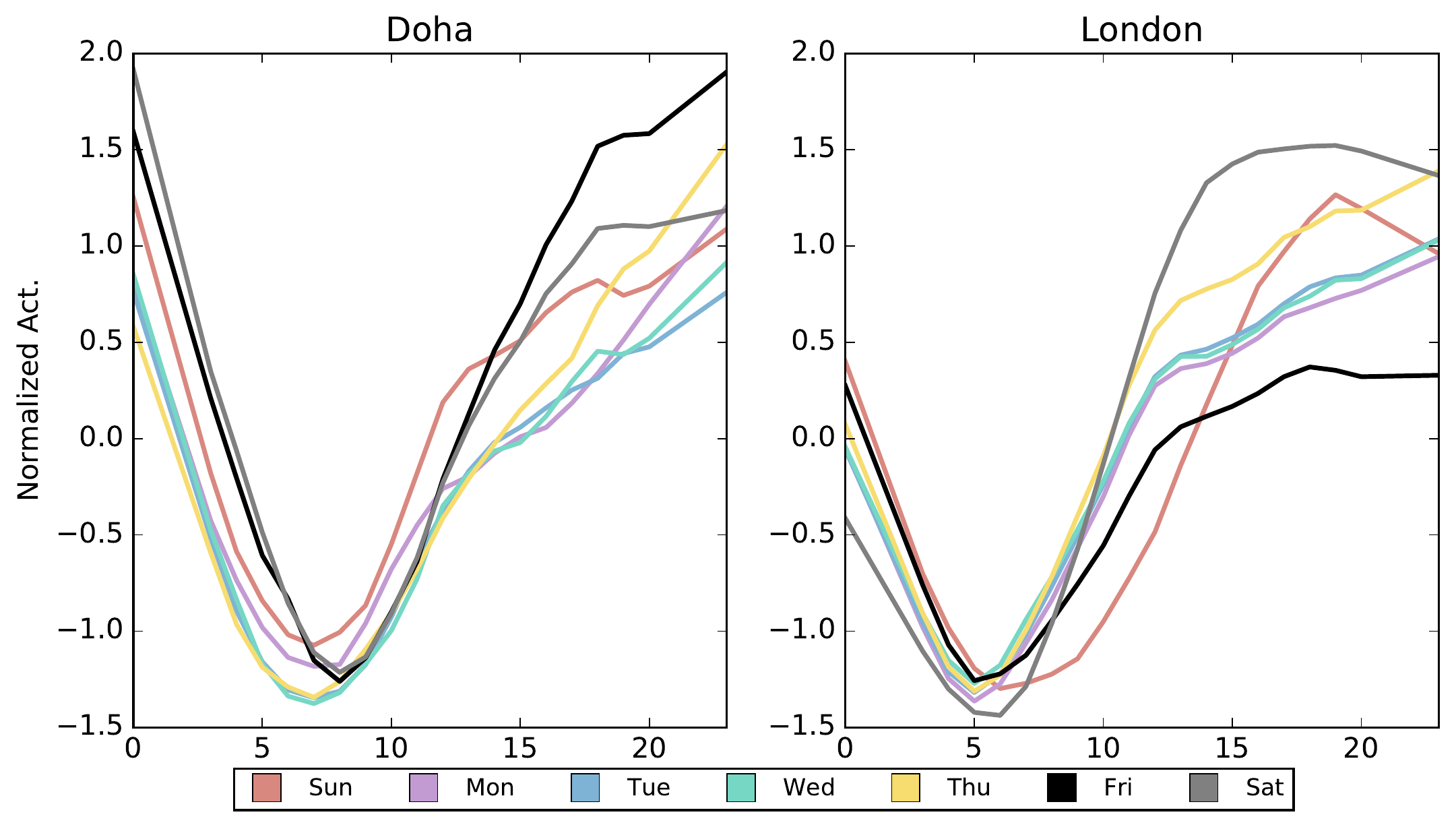}
\caption{Time-series of the typical daily signatures of Doha (left) and London (right). $x-axis$ represents the hours of the day, starting from 00:00 (midnight) through 23:00.}
\label{fig:typical_7days}
\end{figure}

The typical daily signatures shown in Figure~\ref{fig:typical_7days} reveal some interesting patterns. For instance, the city of London wake-up slowly on Sundays, probably due the intense and late activity observed on Saturday nights. On the contrary, Sundays are the day where the city of Doha wake-up quickly compared to other days, which is due to the fact that the week starts on Sunday in Qatar.

\subsection{Insights at the Micro Level of Cities}
In this section, we compare the typical signature of different neighborhoods within the same city to see to which extend social media can unveil differences in local dynamics. Similar kind of analysis has been conducted in the past using proprietary mobile data for the cities of Rome~\cite{Reades:2007} and London~\cite{Grauwin2015}. Thus, we focus in the following on the city of Doha. We picked 4 areas that are dominated by different types of activities: \textit{\bf Souq Wakif}, which is the most visited site in Doha that gathers restaurants, coffee shops, and gift shops. \textit{\bf West Bay}, which is the most westernized neighborhood in the city with many sky-scrappers and hotels. \textit{\bf Industrial Area}, which is the place where lots of industries are based as well as car workshops. \textit{\bf Hamad International Airport (HIA)} which as its name indicates is the only international airport in Doha.

\begin{figure}[!h]
\centering
\includegraphics[width=0.79\columnwidth]{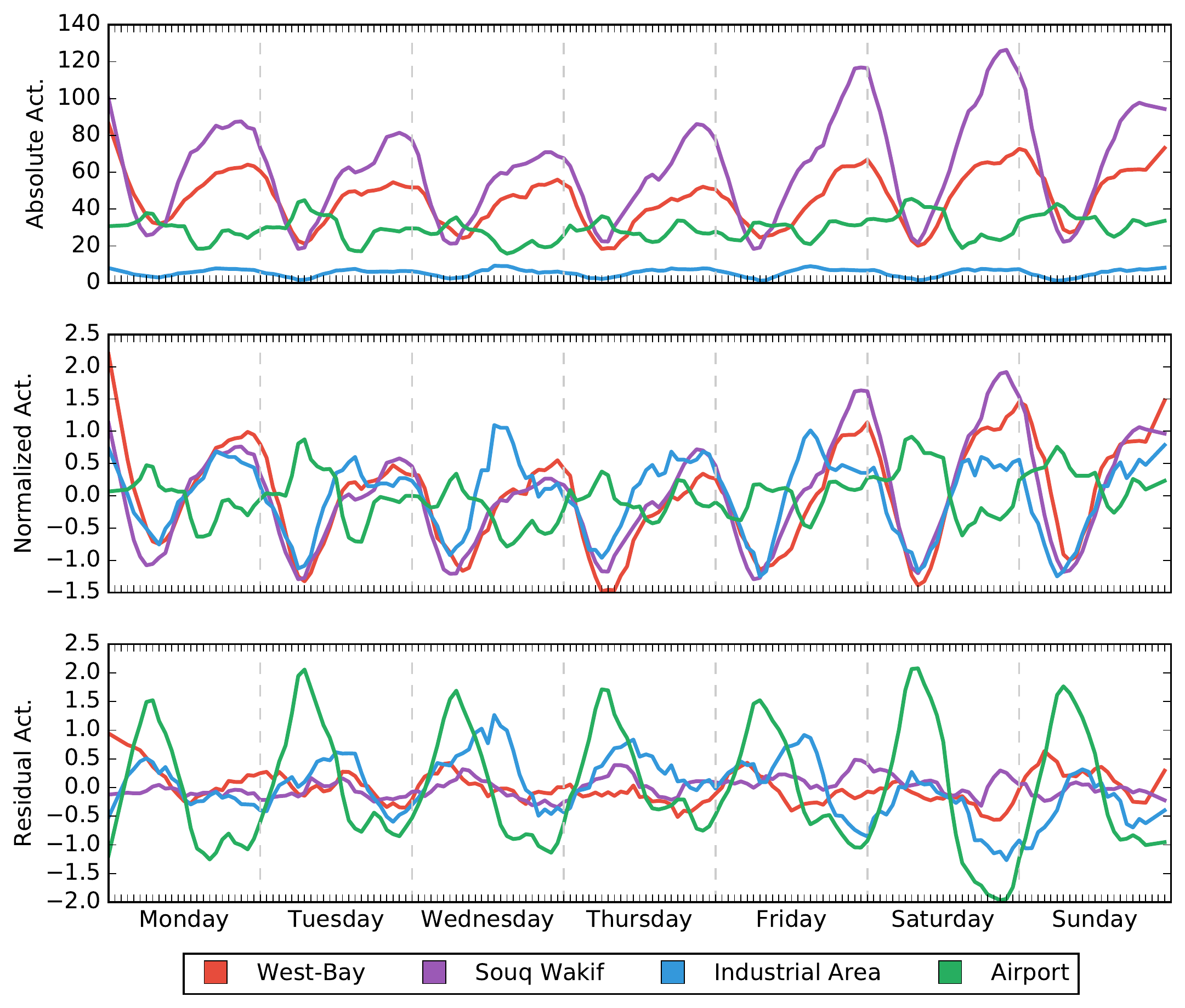}
\caption{Typical weekly signatures of the four selected landmark areas in Doha.}
\label{fig:doha_landmarks}
\end{figure}

Figure~\ref{fig:doha_landmarks} plots the TWS time-series for the four areas selected in Doha. A simple eyeballing allows to see the important difference in term of absolute activity (top panel). As expected, Souq Wakif, which is the most visited and most vibrant site in Doha acquires the highest activity, followed by West-Bay, and the airport. The Industrial Area, which is mainly populated by workers, shows lower level of activity. In order to allow better analysis, we computed again the residual time-series for the four areas ($TWS_{area}^{res}$), with a slightly different definition: for each hour of the week, we subtract the average normalized activity of the whole city from the normalized activity of the area. More formally, 
$TWS_{area}^{res}(t) = TWS_{area}^{norm}(t) - TWS_{city}^{norm}(t).$

The resulting time-series are plotted in bottom panel in Figure~\ref{fig:doha_landmarks}. We can see that activity in industrial area always drops around 4pm, which correspond to the time most workshops close. The airport shows an interesting pattern in which activity is mainly happening early in the morning (the growth starts around mid-night) and spikes around 10:00 in the morning, before it starts fading out throughout the day. This shifted rhythm can be due to people on transit who spend the night at the airport and who tend to be more active on social media at that time. 
These cues are interesting in that they back our assumption that social media footprint of users is a good enough data source to capture different rhythms and patterns that one expect to exist within the same city.

\section{Cluster Analysis}
\label{sec:clustering}

In the previous section, we have demonstrated that using relatively easy and straightforward time-series analysis techniques allows to identify and visualize different urban rhythms happening in different cities or in different areas within the same city. In many cases, the identified patterns matched our a-priori knowledge about the spatio-temporal dynamics of cities such as their overall circadian rhythms, i.e., morning rise of activity, late morning peak, evening drop, etc.  
In the following, we propose to validate our hypothesis with a more principled methodology. The idea here is to group the typical weekly signatures (TWS) of different areas based on their similarity and then map them back into the urban space.
We first recall the TWS representation we use to characterize the urban dynamics associated with different geographic neighborhoods. Next, we introduce $k-means$ clustering algorithm as well as the distance used to measure similarity between time-series. Follows a brief discussion about the method used to find the best number of clusters in the two cities before we discuss our findings.    

\subsection{Time-Series Representation of Urban Activity}
We first need to represent each geographic partition with a time-series reflecting its urban rhythm. Thus, we borrow the technique first introduced in \cite{Reades:2007} and used later in \cite{toole2012inferring,Grauwin2015} to infer the typical weekly signature of different administrative divisions. The difference is that our time-series are built upon social media activity whereas all previous work have used proprietary mobile phone data activity such as calls and sms.
Note that geographic partition can be either well defined administrative divisions such as zones, boroughs, or districts, or simply cells of a created grid of uniform squares. 
In our case, we are interested in clustering administrative divisions (boroughs for London, zones for Doha). 
Thus, we map each social media post into its corresponding area by computing the intersection between the point coordinates (longitude, latitude) and the polygon representing the division which comes in the format of a sequence of points (longitude, latitude) that starts and ends at the same point. Next, we build typical weekly signatures (TWS) for each division. Recall that these time-series are 168 long (= 24 hours $\times$ 7 days), which corresponds to the number of hours in the week. 

\subsection{K-Means}
The goal of clustering is to identify structure in an unlabeled data set by objectively organizing data into homogeneous groups where the within-group-object similarity is minimized and the between-group-object dissimilarity is maximized. 
Hence, clustering is the general problem of partitioning $n$ observations into $k$ clusters, where a cluster is characterized with the notions of homogeneity. Even though many clustering criteria to capture homogeneity and separation have been proposed, the minimum within-cluster sum of squared distances is most commonly used as it expresses both of them. 
Because finding a global optimum is difficult, heuristics such as the $k-means$ method are often used to find a local optimum~\cite{kmeans}.
The general scheme of $k-means$ is described as follows: First, randomly select $k$ items and consider them as cluster seeds. Second, use an iterative procedure that performs two steps in every iteration: (i) assignment step in which each item is assigned to the cluster of its nearest centroid, which is determined with the use of a distance function; (ii) reevaluation of cluster centroids to reflect the changes in cluster memberships. Finally, the algorithm converges either when there is no change in cluster memberships or when the maximum number of iterations is reached. Given that items in our case are TWS time-series, it is important to use a distance function that better captures similarities between then. The next sub-section describes the one that we have used.

\begin{figure*}
\centering
\begin{minipage}{0.39\textwidth}
\subfigure[Doha]{ \includegraphics[width=0.95\columnwidth]{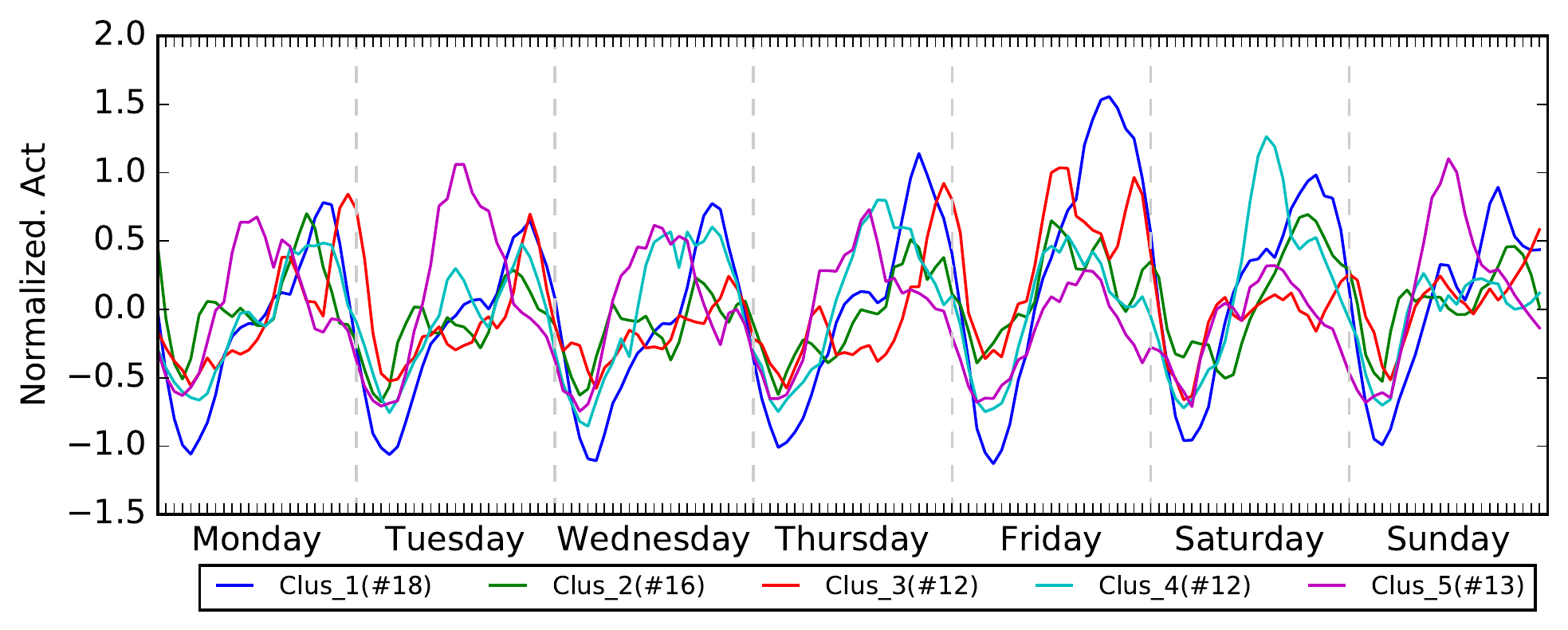}}
\subfigure[London]{ \includegraphics[width=0.95\columnwidth]{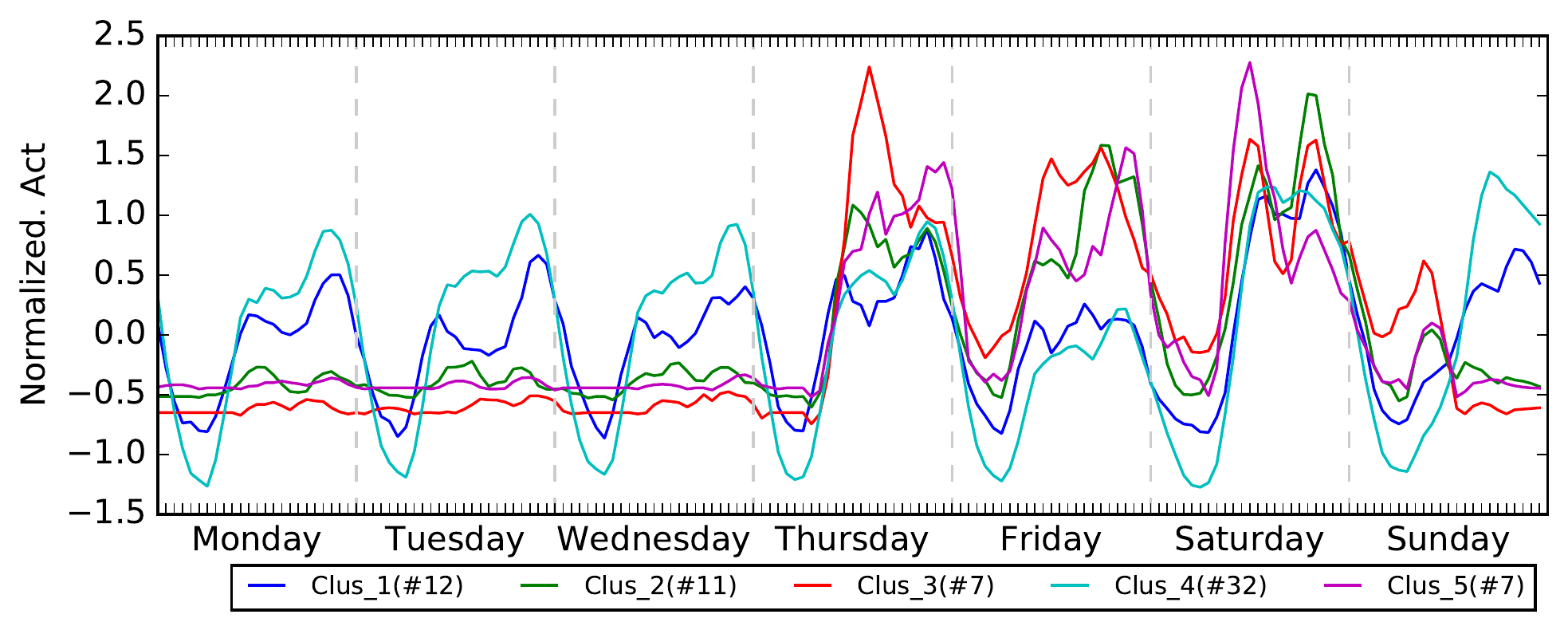}}
\end{minipage}%
\begin{minipage}{0.6\textwidth}
\subfigure[Doha]{\includegraphics[width=0.37\columnwidth]{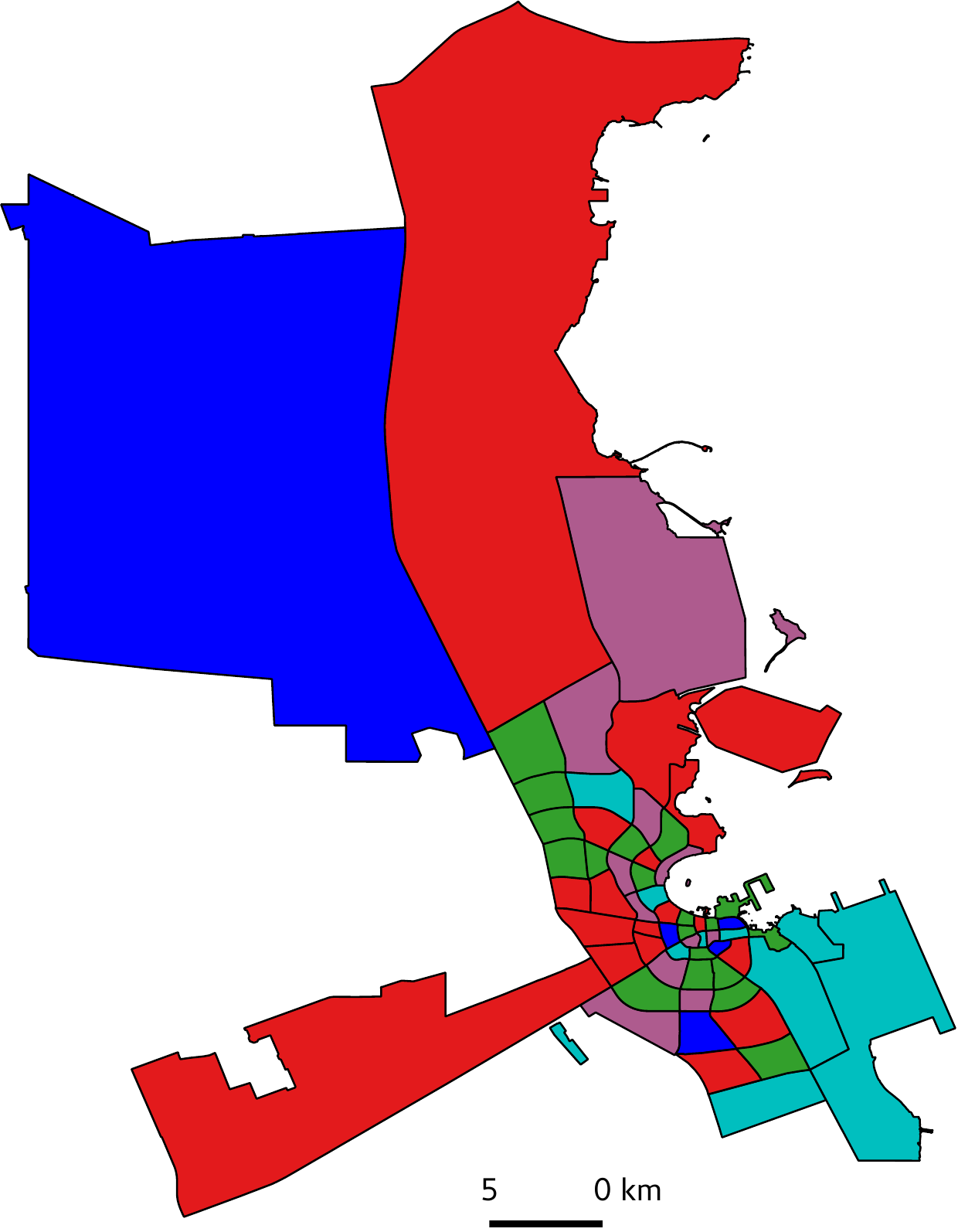}}
\hfill
\subfigure[London]{\includegraphics[width=0.64\columnwidth]{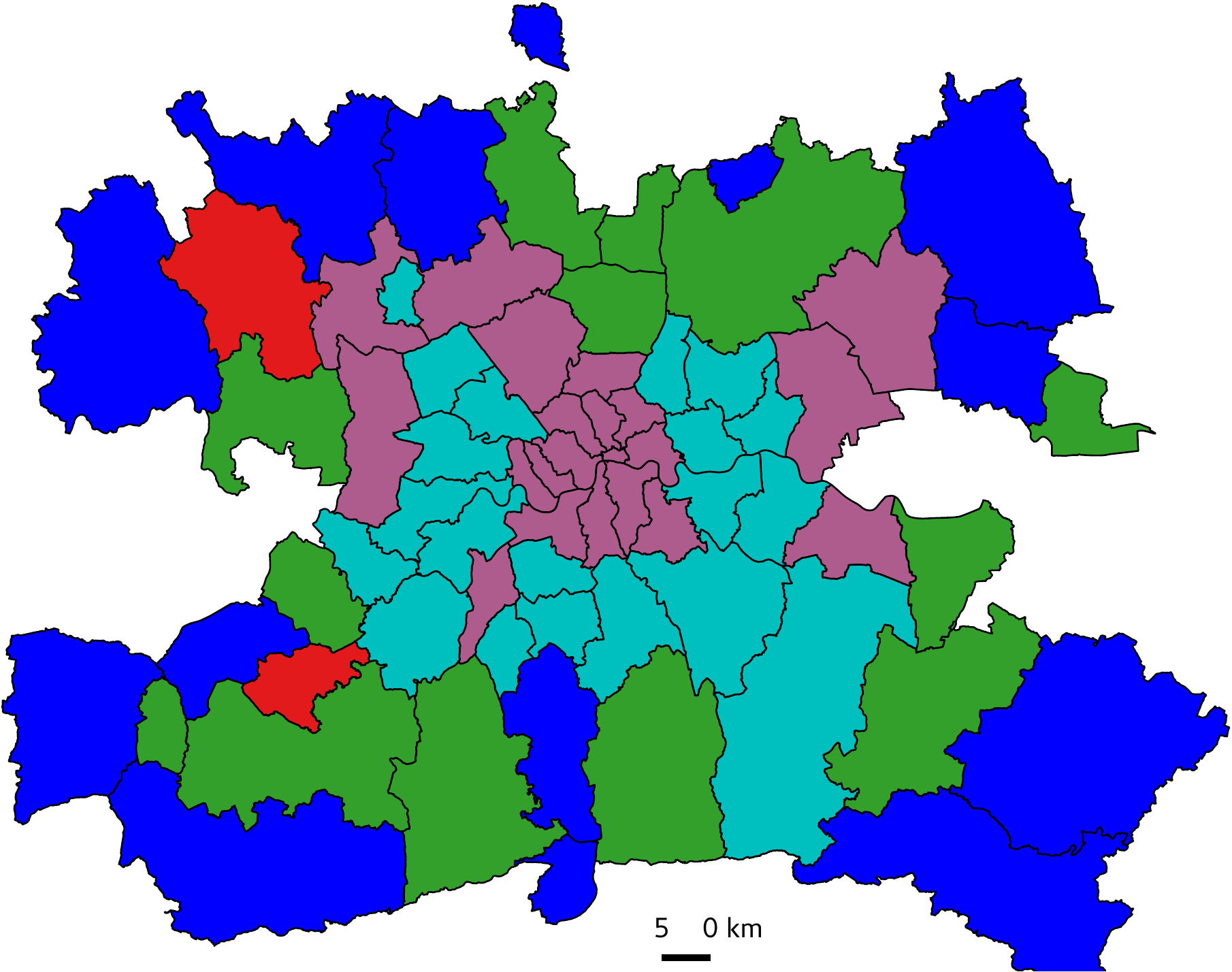}}
\end{minipage}%
\caption{Results of the independent cluster analysis. Panels (a) and (b) plot the representative (average) TWS time-series for the different clusters found in Doha and London respectively. Panels (c) and (d) project the clusters into the map of the two cities.}
\label{fig:cluster_analysis}
\end{figure*}

\subsection{Time-Series Distance Measure}
For any clustering algorithm, we need to define the distance function to use in order to assess how similar are two items (time series in our case.) 
The two state-of-the-art steps for time-series comparison are to first $z$-normalize the sequences and then use a distance measure to determine their similarity.
The most widely used distance metric is Euclidean distance that compares time-series of the same length by computing the root square of the sum of squared element-wise subtractions. However, despite of its simplicity, Euclidean distance shows several shortcomings related to time-series invariances such as scaling, translation, and shifting \cite{dtw,wang2013experimental,keogh2005exact}.
Another popular distance measure is Dynamic Time Warping (DTW)\cite{dtw}. DTW can be seen as an extension of Euclidean distance that offers a local (non-linear) alignment. To achieve that, an $m$-by-$m$ matrix $M$ is constructed, with the Euclidean distance between any two points of the two time-series. A warping path $W = {w_1, w_2, \ldots, w_k}$, where $k \geq m$, is a contiguous set of matrix elements that defines a mapping between the two time-series $\vec{x}, \vec{y}$: $DTW(\vec{x}, \vec{y}) =  min\sqrt{\sum_{i=1}^{k} w_i}$. DTW is shown to be more appropriate for many time-series tasks. For instance, Wang et al.\cite{wang2013experimental} extensively evaluated 9 distance measures and several variants thereof; They found that DTW along with  some of its variants perform exceptionally well in comparison to other measures. In our study, we use an optimized version of DTW known as Keogh Lower-Bound (LB-Keogh) approximation \cite{keogh2005exact} to efficiently and effectively evaluate the similarity of time-series. 

\subsection{Find the optimum number of clusters K}
\begin{figure}
  \centering
  \includegraphics[width=0.5\columnwidth]{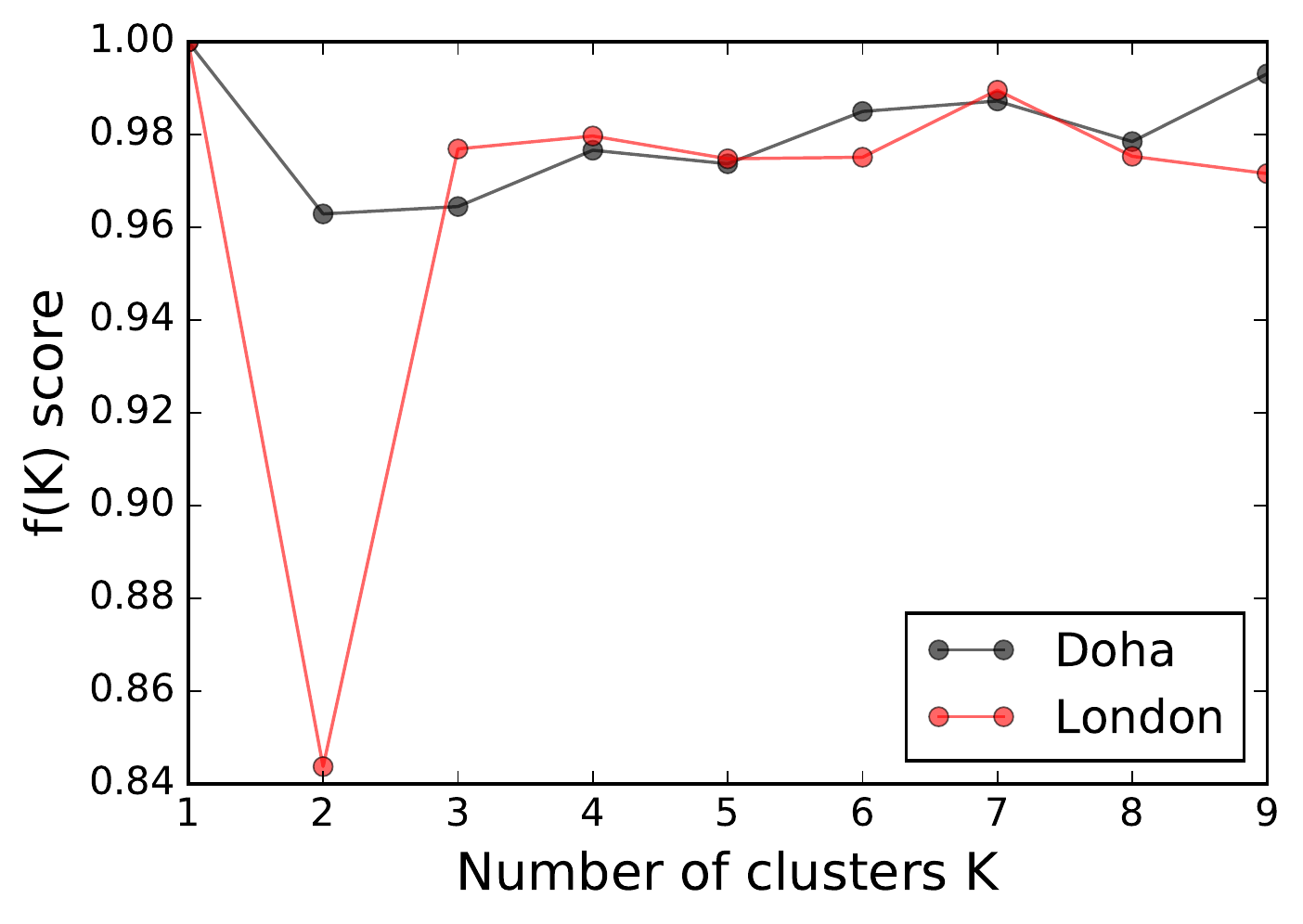}
  \caption{Using $f-scores$ to determine the best value(s) of $K$. Best values coincide with local minima of $f(K)$}
  \label{fig:bestK}
\end{figure}

Finding the optimal number of clusters is a fundamental problem in $k-means$ clustering and alike partitioning methods which requires the number of clusters to be provided as an input. Unfortunately, there is no definitive method to determine this optimal number of clusters which usually depends on the distance function used for measuring similarities between items in the data set, as well as the distribution of the items. Despite the existence of several heuristics that researchers use, including direct methods in which we seek to optimize for a criterion such as the sums of squares (e.g.,  elbow~\cite{elbow:1996}, Silhouette scores~\cite{silhouettes:1987}) and statistical testing that proceed by comparing evidence against null hypothesis (e.g., gap statistic~\cite{tibshirani2001estimating,pham2005selection}), there is no real consensus on which one to use. Indeed, the choice of $k$ is often guided by domain expertise, intuition, or the level of resolution desired. 
In our case, we used the optimized version of gap static method proposed in~\cite{pham2005selection} to guide our choice of $K$. For both cities, we found $K=2$ to be a good choice as suggested in Figure~\ref{fig:bestK}. However, other possible good values of $K$ are 5 and 8. Obviously, a small value of $K$ favors the creation of large and broadly similar clusters whereas a high value of $K$ favors the creation of smaller yet specific clusters. Thus, we set $K$ to be 5 as a good  compromise between specificity and generality.

Figure~\ref{fig:bestK} shows the optimal number of clusters for both Doha and London.  

\begin{figure}
\centering
\subfigure[Doha (2 clusters)]{\includegraphics[width=0.3\columnwidth]{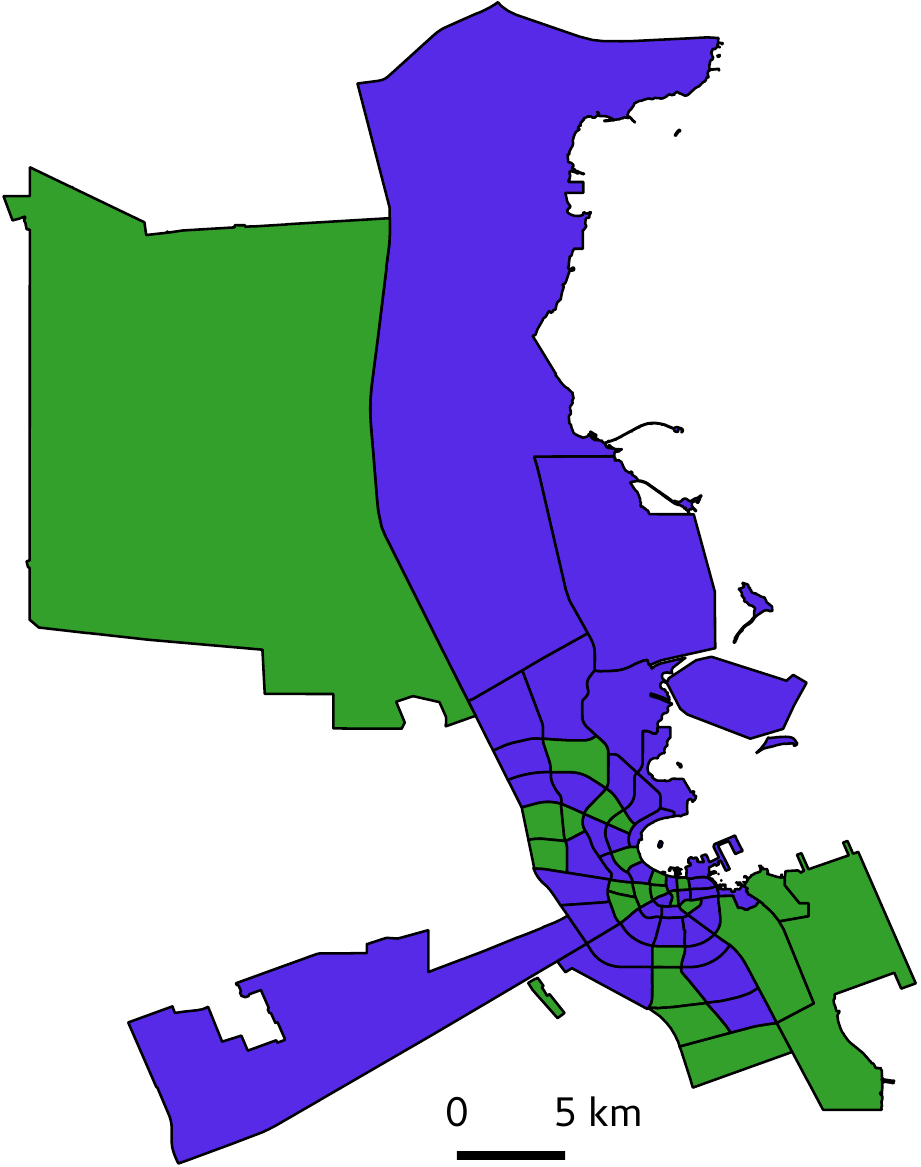}}
\subfigure[London (2 clusters)]{ \includegraphics[width=0.52\columnwidth]{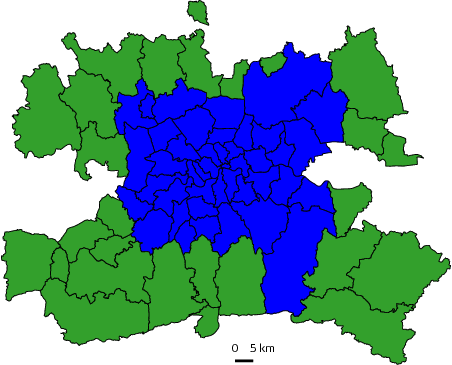}}
\subfigure[Doha (5 clusters)]{\includegraphics[width=0.3\columnwidth]{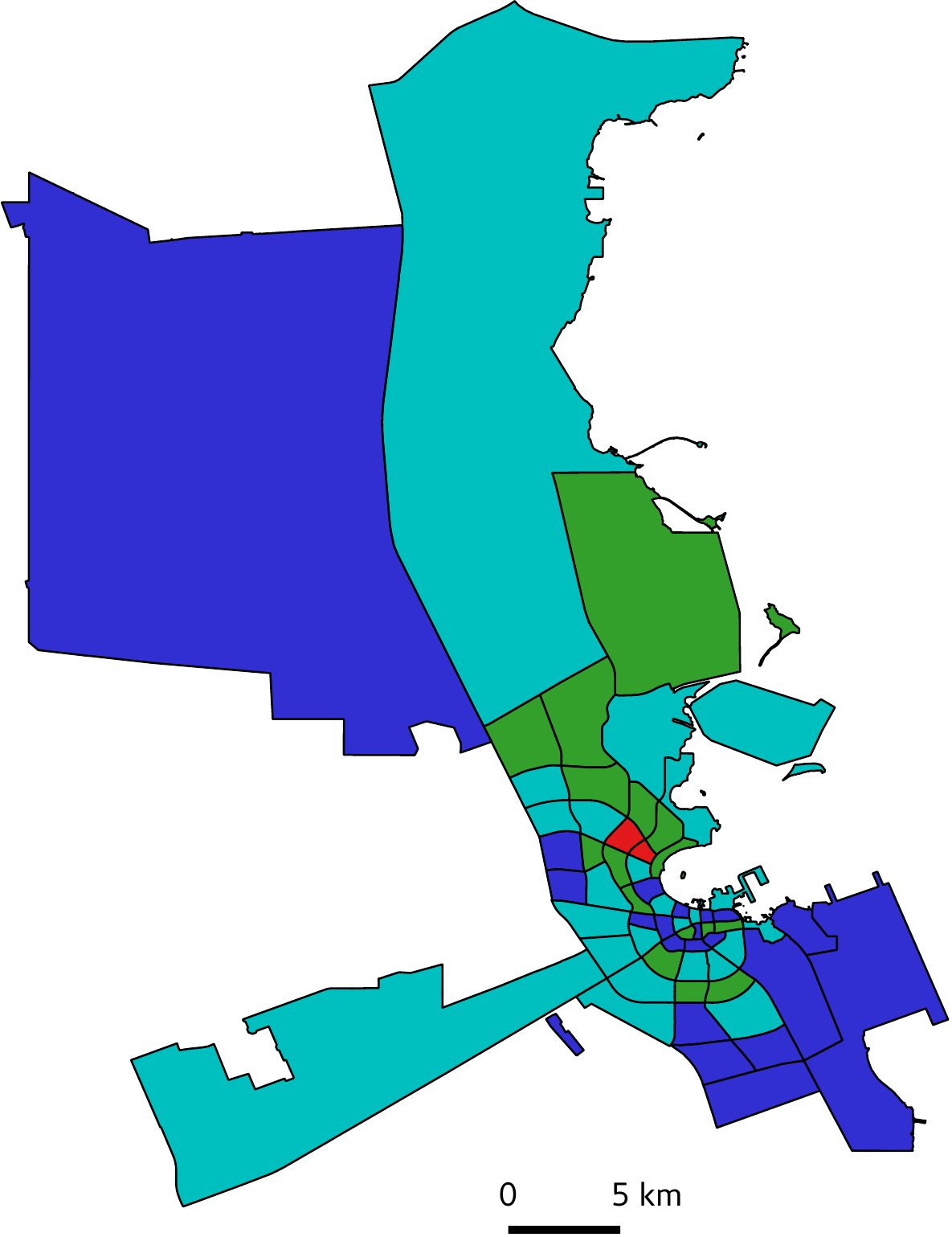}}
\subfigure[London (5 clusters)]{ \includegraphics[width=0.52\columnwidth]{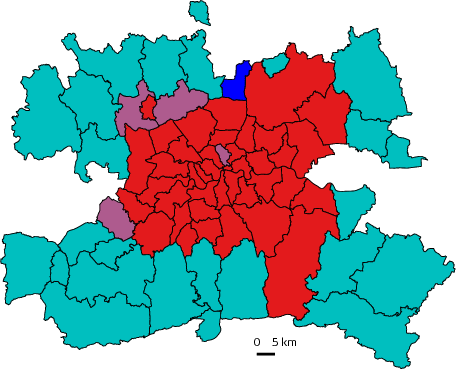}}
\subfigure[Doha (8 clusters)]{\includegraphics[width=0.3\columnwidth]{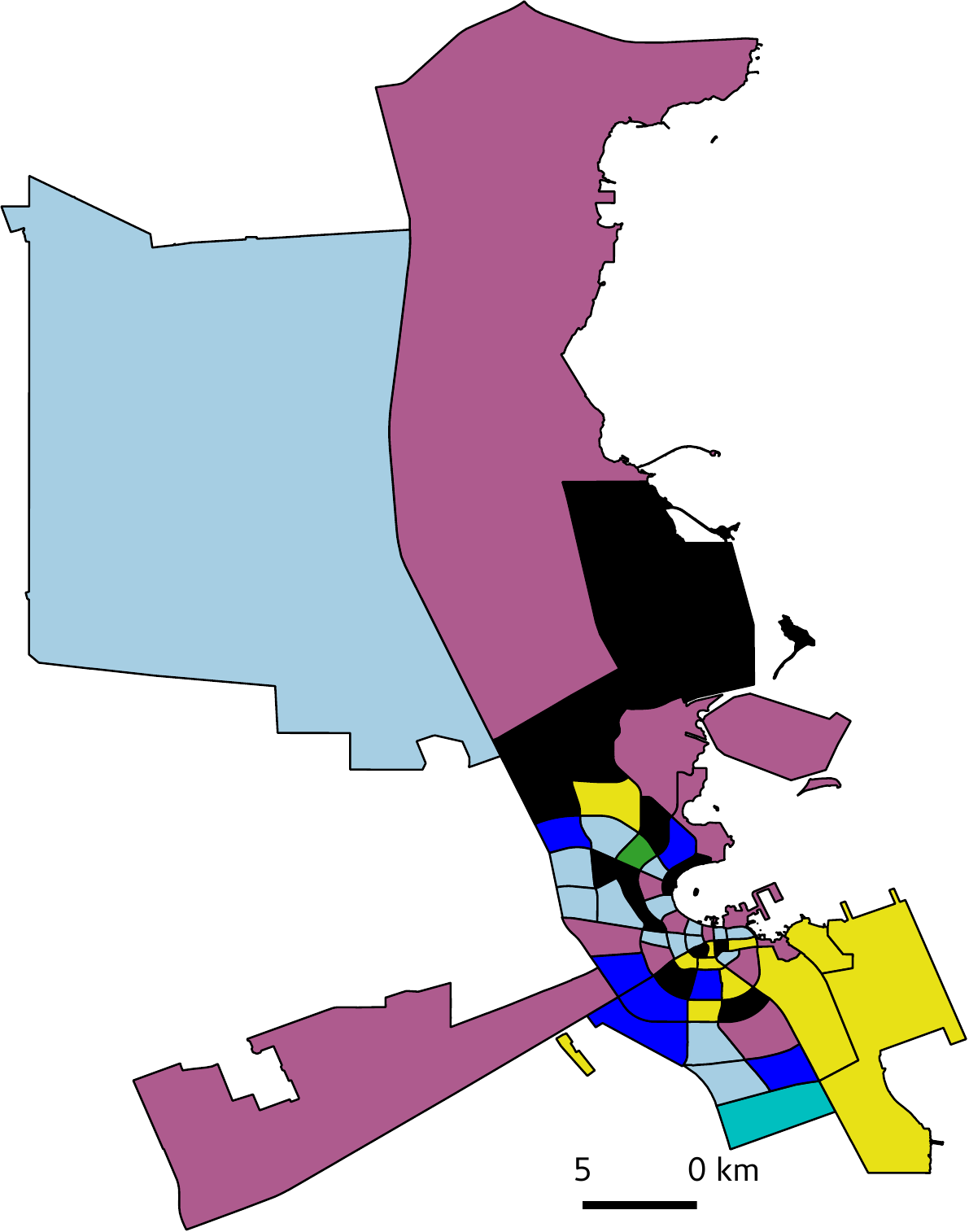}}
\subfigure[London (8 clusters)]{ \includegraphics[width=0.52\columnwidth]{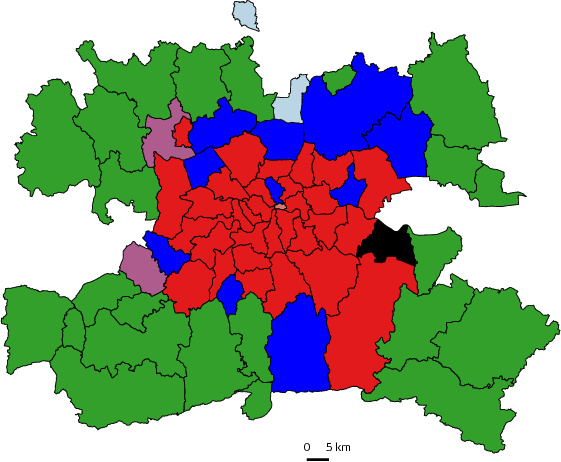}}
\caption{Clustering Doha zones and London boroughs all together with different values of $K$ (number of clusters.) For each $K$, divisions from the two cities sharing the same color have similar TWS time-series.}
\label{fig:transveral_kmeans}
\end{figure}

\subsection{Discussion}
We first requested clustering the administrative divisions of each city into five mutually exclusive partitions. This is done via independent k-means clustering analysis for each city. 
In panels (a) and (b) of Figure~\ref{fig:cluster_analysis}, we plot the normalized TWS time-series of each cluster centroid found in Doha and London respectively. A close inspection of the Doha's TWS time-series reveals the presence different dynamic patterns taking place in different areas of the city. 
For instance, zones belonging to cluster $\#5$ (purple) are characterized by an activity pattern that builds up in the early morning, maximizes around noon, and starts fading out in the afternoon to completely die around midnight. This pattern is typical of areas dominated by day activity facilities such as administrations and schools. For the record, public schools terminate classes at 01:30pm in Qatar, day work in public administration ends at 03:30pm.
Zones belonging to cluster $\#1$ (blue) on the contrary are characterized by a later spike happening around 08:00pm. In fact, the corresponding TWS time-series of this cluster shows a slow yet important rise of activity that starts around 06:00am and keeps building up until it spikes around 08:00pm, then fades out. The important level of late activity is a good indicator of the presence of night life spots (e.g., restaurants, bars, etc.) in the corresponding zones. 
Zones in cluster $\#3$ show a distinguishable daily bimodal distribution of activity. The first small peak usually occurs before noon whereas the second major peak occurs around midnight. \\
London's TWS time-series on the other hand also show some interesting insights about the different urban rhythms that characterize different boroughs. The most striking one is related to clusters $\#1$ (blue) and $\#4$ (azure) which have two activity phases: A steady phase recorded on Sunday, Monday, Tuesday, and Wednesday, and a more vibrating phase recorded on Thursday, Friday and Saturday. 
While the remaining clusters show relatively similar patterns, they greatly differ from each other when we looked into their absolute volume of weekly activity.\\
Projecting TWS time-series of clusters into the map reveals some geographical grouping of zones belonging to similar clusters. Panels (c) and (d) in Figure~\ref{fig:cluster_analysis} show the result of this projection. The obtained clusters for London demonstrate an interesting geographical structure. That is, boroughs on cluster $\#5$ (purple) are concentrated toward the center of the city. These boroughs are surrounded by another circle of boroughs belonging to cluster $\#4$ (azure). Then come clusters $\#1$ (blue) and $\#2$ (green) in the outer circle of the city. Doha's clusters are organized into a less obvious spatial configuration that makes it difficult to interpret via simple visual inspection.\\     
So far, we saw that social media data, and Twitter in particular, can be used to reveal different dynamics and patterns grouped into clusters that share similar typical signatures. In order to see the extent to which the two cities are similar, we run a transversal k-means clustering algorithm on both cities of Doha and London at once. To do so, we put all administrative divisions (71 Doha zones and 69 London boroughs) into one bag in which each division is characterized with its normalized typical weekly signature (TWS). Next, we request k-means to digests all divisions together and partition them into clusters of divisions with similar signatures. Figure~\ref{fig:transveral_kmeans} plots the projection of the clusters into the maps of the two cities, for different values of $K \in \{2, 5, 8\}$. Divisions with the same color in the two cities belong to the same cluster, and thus share similar typical signatures.

There are few interesting observations that we can make here. First, in the case of $K=2$, we clearly see that central boroughs of London and central zones of Doha are grouped into the same cluster (blue), whereas peripheral boroughs and zones are grouped into another cluster (green). What is happening here is that k-means is mainly partitioning divisions based on level of activities. For the case of $K=5$, the two cities have shown different distributions of clusters (see Figure~\ref{fig:transveral_kmeans} (c) and (d)). Now, the mainly common pattern of the two cities in the one concerning boroughs at the periphery of London and zones with low activity in Doha (represented with azure color.) The red cluster seems to be specific to center London, with only one zone in Doha belonging to the same cluster. Note that this zone is close the the central business district of Doha. In contrast, green and blue clusters seem to be specific to Doha. Finally, in the case of $K=8$ (see Figure~\ref{fig:transveral_kmeans} (c) and (d)), we see that k-means picked on more specificities for the two cities, with very little transversal clusters.

\section{Conclusion}
\label{sec:conclusion}

In this paper, we investigated the use of user generated social media posts, in particular geo-located Twitter data, for modeling the urban dynamics of two cities: Doha (Qatar) and London (UK.)
Following some previous models that used mobile phone data, we characterized different cities and neighborhoods using typical weekly signatures (TWS), which are time-series reflecting the average volume of social-media activity witnessed in different hours and days of the week. Initial data analysis enabled the comparaison between the circadian rhythms of the two cities in which people typically wake-up in the morning, commute to work, and go back home in late in afternoon. We could also spot the change in dynamics that takes place during weekends. In order to closely inspect the commonalities and differences of fine-grained neighborhoods, we conducted a cluster analysis in which similar areas are bundled together. The cluster analysis revealed that close neighborhoods tend to share similar rhythms.

In the future, we will work on integrating more data sources such as distributions of land use and points-of-interest in order to better related the typical weekly signatures to human activities. This includes for instance trying to infer the particular land use of a parcel from its typical signature. We want also to exploit the content of social media posts, such as text, urls, and photos, to create more accurate and richer neighborhood profiles.

\bibliographystyle{splncs03}
\bibliography{biblio}

\end{document}